\documentstyle[amsfonts,12pt]{article}

\title{Existence of Dyons in Minimally Gauged\\ Skyrme Model via Constrained Minimization}
\author{Zhifeng Gao\footnote{Email address: gzf@henu.edu.cn}
\\Institute of Contemporary Mathematics\\School of Mathematics\\Henan University\\
Kaifeng, Henan 475001, PR China\\ \\ Yisong Yang\footnote{Email address: yyang@math.poly.edu}
\\Department of Mathematics\\Polytechnic
Institute of New York University\\Brooklyn, New York 11201, USA
}
\date{}
\newcommand{\bfR}{{\Bbb R}}

\newcommand{\bfZ}{{\Bbb Z}}
\newtheorem{oldtheorem}{Theorem}[section]
\newtheorem{oldassertion}[oldtheorem]{Assertion}
\newtheorem{oldproposition}[oldtheorem]{Proposition}

\newtheorem{oldlemma}[oldtheorem]{Lemma}
\newtheorem{olddefinition}[oldtheorem]{Definition}
\newtheorem{oldclaim}[oldtheorem]{Claim}
\newtheorem{oldcorollary}[oldtheorem]{Corollary}

\newenvironment{theorem}{\begin{oldtheorem}$\!\!\!${\bf.}}{\end{oldtheorem}}

\newenvironment{lemma}{\begin{oldlemma}$\!\!\!${\bf.}}{\end{oldlemma}}

\newbox\qedbox
\newenvironment{proof}{\smallskip\noindent{\bf Proof.}\hskip \labelsep}%
                        {\hfill\penalty10000\copy\qedbox\par\medskip}

\newcommand{\dd}{\mbox{d}}
\newcommand{\ee}{\end{equation}}
\newcommand{\be}{\begin{equation}}\newcommand{\bea}{\begin{eqnarray}}
\newcommand{\eea}{\end{eqnarray}}
\newcommand{\e}{\mbox{e}}
\newcommand{\pa}{\partial}\newcommand{\Om}{\Omega}
\newcommand{\vep}{\varepsilon}

\newcommand{\nn}{\nonumber}

\newcommand{\lm}{\lambda}

\begin{document}
\maketitle
\begin{abstract}
We prove the existence of electrically and magnetically charged particle-like static solutions, known as dyons, in the minimally gauged
Skyrme model developed by Brihaye, Hartmann, and Tchrakian. The solutions are spherically symmetric, depend on two continuous parameters,
and carry unit monopole and magnetic charges but continuous Skyrme charge and non-quantized electric charge induced from the 't Hooft electromagnetism.
The problem amounts to obtaining a finite-energy critical point of an indefinite action functional, arising from the presence of electricity and the Minkowski spacetime signature.
The difficulty with the absence of the Higgs field is overcome by achieving suitable strong convergence and obtaining uniform decay estimates at singular boundary points
so that the negative sector of the action functional becomes tractable.

\medskip

\end{abstract}

\section{Introduction}
\setcounter{equation}{0}

It was Dirac \cite{D} who first explored the electromagnetic duality in the Maxwell equations and came up with a mathematical
formalism of magnetic monopoles, which was initially conceptualized by P. Curie \cite{Curie}. Motivated by the search of a quark model,
Schwinger \cite{S} extended the study of Dirac \cite{D} to obtain a new class of particle-like solutions of the Maxwell equations carrying both electric and
magnetic charges, called dyons, and derived an elegant charge-quantization formula for dyons, generalizing that of Dirac for monopoles. However, both the Dirac monopoles and Schwinger dyons
are of infinite energy and deemed unphysical. In the seminal works of Polyakov \cite{Po} and 't Hooft \cite{'t}, finite-energy smooth monopole solutions were obtained
in non-Abelian gauge field theory. Later, Julia and Zee \cite{JZ} extended the works of Polyakov and 't Hooft and obtained finite-energy smooth dyon solutions in the same
non-Abelian gauge field theory framework. See Manton and Sutcliffe \cite{MS} for a review of monopoles and dyons in the context of a research monograph on topological solitons.
See also \cite{A,Go,R} for some earlier reviews on the subject.
In contemporary physics, monopoles and dyons are relevant theoretical constructs for an interpretation of quark confinement \cite{Gr,Man,SY}.

Mathematically, the existence of monopole and dyons is a sophisticated and highly challenging problem. In fact, the construction of monopoles and dyons was first made possible 
in the critical Bogomol'nyi \cite{B} and Prasad--Sommerfeld \cite{PS} (BPS) limit, although an analytic proof of existence of spherically symmetric unit-charge monopoles was also obtained
roughly at the same time \cite{BPST}. A few years later, the BPS monopoles of multiple charges were obtained by Taubes \cite{JT,Taubes} using a gluing technique to patch a
distribution of widely separated unit-charge BPS monopoles together. Technically, the existence of dyons is a more difficult problem even for spherically symmetric solutions of unit charges.
The reason is that the presence of electricity requires a non-vanishing temporal component of gauge field as a consequence of the 't Hooft construction \cite{tH} of electromagnetism so that
the action functional governing the equations of motion becomes indefinite due to the Minkowski spacetime signature. In fact, the original derivation of the BPS dyons is based on
an internal-space rotation of the BPS monopoles, also called the Julia--Zee correspondence \cite{A}. An analytic proof for the existence of the Julia--Zee dyons \cite{JZ},
away from the BPS limit, was obtained by
Schechter and Weder \cite{SW} using a constrained minimization method. Developing this method, existence theorems have been established for dyons in the Weinberg--Salam electroweak
theory \cite{CM,Yw,Ybook}, 
and in the Georgi--Glashow--Skyrme model \cite{BKT,LY}, as well as for the Chern--Simons vortex equations \cite{CGSY,LPY}.

It is well known that the Skyrme model \cite{S1,S2} is important for baryon physics \cite{GP,GM,MRS,ZB} and soliton-like solutions in the Skyrme model, called Skyrmions, are used to model
elementary particles. Thus, in order to investigate inter-particle forces among Skyrmions, gauge fields have been introduced into the formalism
 \cite{AR,BHT,BKT,CW,DF,Eilam,PT,W1}. Here, we are interested in the minimally gauged Skyrme model studied by Brihaye, Hartmann, and Tchrakian \cite{BHT}, where the Skyrme (baryon) charge may be
prescribed explicitly in a continuous interval. The Skyrme map is hedgehog and the presence of gauge fields makes the static solutions carry both electric and magnetic charges. In other words,
these gauged Skyrmions are dyons. In \cite{BHT}, numerical solutions are obtained which convincingly support the existence of such solutions. The purpose of this paper is to give an analytic
proof for the existence of these solutions, extending the methods developed in the earlier studies \cite{LY,SW,Yw,Ybook} for the dyon solutions in other models in field theory described above.
See also \cite{BR,FR}.
Note that, since here we are interested in the minimally gauged Skyrme model where no Higgs field is present, we lose the control over the negative terms
in the indefinite action functional which can otherwise be controlled if a Higgs field is present \cite{BKT,LY,SW,Yw,Ybook}. In order to overcome this difficulty, we need to obtain suitable
uniform estimates for a minimizing sequence at singular boundary points and to achieve strong convergence results, for the sequences of the negative terms.

The contents of the rest of the paper are outlined as follows. In Section 2, we review the minimally gauged Skyrme model of Brihaye--Hartmann--Tchrakian \cite{BHT} and then state our main
existence theorem for dyon solutions. It is interesting that the solutions obtained are of unit monopole and magnetic charges but continuous Skyrme charge and non-quantized electric charge.
In the subsequent three sections, we establish this existence theorem.
In Section 3, we prove the existence of a finite-energy critical point of the indefinite action functional by formulating and solving a constrained minimization problem.
In Section 4, we show that the critical point obtained in the previous section for the constrained minimization problem solves the original equations of motion by proving that
the constraint does not give rise to a Lagrange multiplier problem. In Section 5, we study the properties of the solutions. In particular, we obtain some uniform decay
estimates which allow us to describe the dependence of the ('t Hooft) electric charge on the asymptotic value of the electric potential function at infinity.
\section{Dyons in the minimally gauged Skyrme model}
\setcounter{equation}{0}

As in the classical Skyrme model \cite{S1,S2}, the minimally gauged Skyrme model of Brihaye--Hartmann--Tchrakian \cite{BHT} is built around a wave map, $\phi=(\phi^a)$ ($a=1,2,3,4$), from
the Minkowski spacetime $\bfR^{3,1}$ of signature $(+---)$ into the unit sphere, $S^3$, in $\bfR^4$, so that $\phi$ is subject to the constraint $|\phi|^2=(\phi^a)^2=1$, where and in the sequel,
summation convention is implemented over repeated indices. Finite-energy condition implies that $\phi$ approaches a fixed vector in $S^3$, at spatial infinity. Thus,
at any time $t=x^0$, $\phi$ may be viewed as a map from $S^3$, which is a one-point compactification of $\bfR^3$, into $S^3$. Hence, $\phi$ is naturally characterized by an integral class,
say $[\phi]$, in
the homotopy group $\pi_3(S^3)=\bfZ$. The integer $[\phi]$, also identified as the Brouwer degree of $\phi$, may be represented as a volume integral of the form
\be \label{B}
B_\phi=[\phi]=\frac1{12\pi^2}\int_{\bfR^3}\vep_{ijk}\vep^{abcd}\pa_i\phi^a\pa_j\phi^b\pa_k\phi^c\phi^d\,\dd x,
\ee
where $i,j,k=1,2,3$ denote the spatial coordinate indices and $\vep$ is the Kronecker skewsymmetric tensor. 
This topological invariant is also referred to as the Skyrme charge or baryon charge. Let $(\eta_{\mu\nu})=\mbox{diag}\{1,-1,-1,-1\}$ ($\mu,\nu=0,1,2,3$) be
the Minkowski metric tensor and $(\eta^{\mu\nu})$ its inverse.
We use $|A_\mu|^2=\eta^{\mu\nu}A_\mu A_\nu$ to denote the squared Minkowski norm of a 4-vector $A_\mu$ and $A_{[\mu}B_{\nu ]}=A_\mu B_\nu-A_{\nu}B_\mu$ to denote the skewsymmetric tensor product
of $A_\mu$ and $B_\mu$. The Lagrangian action density of the Skyrme model \cite{S1,S2} is of the form
\be 
{\cal L}=\frac12\kappa_1^2|\pa_\mu \phi^a|^2-\frac12 k_2^4|\pa_{[\mu}\phi^a \pa_{\nu ]}\phi^b|^2,
\ee 
where $\kappa_1,\kappa_2>0$ are coupling constants. The model is invariant under any internal space rotation. That is, the model enjoys a global $O(4)$ symmetry. Such a symmetry is broken 
down to $SO(3)$ by
suppressing the vacuum manifold to a fixed point, say ${\bf n}=(0,0,0,1)$, which may be specified by inserting a potential term of the form
\be 
V=\lm(1-\phi^4)=\lm (1-{\bf n}\cdot\phi)^4,\quad\lm>0,
\ee
into the Skyrme Lagrangian density.

The `residual' $SO(3)$ symmetry is now to be gauged. In order to do so, we follow \cite{BHT} to set $\phi=(\phi^a)=(\phi^\alpha,\phi^4)$ and replace the common derivative
by the $SO(3)$ gauge-covariant derivative
\be 
D_\mu\phi^\alpha=\pa_\mu\phi^\alpha+\vep^{\alpha\beta\gamma}A^\beta_\mu\phi^\gamma,\quad\alpha,\beta,\gamma=1,2,3,\quad D_\mu\phi^4=\pa_\mu\phi^4,
\ee
where $A^\alpha_\mu$ is the $\alpha$-component of the $SO(3)$-gauge field ${\bf A}_\mu$ in the standard isovector representation ${\bf A}_\mu=(A_\mu^\alpha)$, which induces the gauge field
strength tensor
\be 
{\bf F}_{\mu\nu}=\pa_\mu {\bf A}_\nu-\pa_\nu{\bf A}_\mu+{\bf A}_\mu\times {\bf A}_\nu=(F^\alpha_{\mu\nu}).
\ee
As a result, the $SO(3)$ gauged Skyrme model is then defined by the Lagrangian density \cite{BHT}
\begin{eqnarray}
\mathcal{L}
&=&-\kappa^4_0|F^\alpha_{\mu\nu}|^2+\frac{1}{2}\kappa^2_1|D_\mu\phi^a|^2-\frac{1}{2}\kappa^4_2|D_{[\mu}\phi^aD_{\nu]}\phi^b|^2-V_\omega(\phi),
\label{L}\end{eqnarray} 
where $\kappa_0>0$ and the potential function $V_\omega$ is taken to be
\be \label{Vo}
V_\omega(\phi)=\lm(\cos\omega-\phi^4)^2,\quad 0\leq\omega\leq\pi,
\ee
with $\omega$ an additional free parameter which is used to generate a rich vacuum manifold defined by
\be \label{V}
|\phi^\alpha|=\sin\omega,\quad \phi^4=\cos\omega.
\ee

In order to stay within the context of minimal coupling, we shall follow \cite{BHT} to set $\lm=0$ to suppress the potential term (\ref{Vo}) but maintain the vacuum manifold (\ref{V}) by imposing
appropriate boundary condition at spatial infinity.

Besides, since the topological integral (\ref{B}) is not gauge-invariant, we need to replace it by the quantity \cite{AT,BHT}
\be 
Q_S=B_{\phi,A}=\frac1{12\pi^2}\int_{\bfR^3}\left(\vep_{ijk}\vep^{abcd}D_i\phi^a D_j\phi^b D_k\phi^c\phi^d-3\vep_{ijk}\phi^4 F_{ij}^\alpha D_k\phi^\alpha\right)\,\dd x,
\ee
as the Skyrme charge or baryon charge. On the other hand, following \cite{Go,Ryder}, the monopole charge $Q_M$ is given by
\be 
Q_M=\frac1{16\pi}\int_{\bfR^3}\vep_{ijk} F^\alpha_{ij}D_k\phi^\alpha\,\dd x,
\ee
which defines the homotopy class of $\phi$ viewed as a map from a 2-sphere near the infinity of $\bfR^3$ into the vacuum manifold described in (\ref{V}) which happens to be a 
2-sphere as well when $\omega\in (0,\pi)$.

Following \cite{BHT}, we will look for solutions under the spherically symmetric ansatz
\bea
A^\alpha_0&=&g(r)\left(\frac{x^\alpha}r\right),\quad
A^\alpha_i=\frac{a(r)-1}{r}\varepsilon_{i\alpha\beta}\left(\frac{x^\beta}r\right),\label{a1}\\
\phi^\alpha&=&\sin f(r)\left(\frac{x^\alpha}r\right),\ \ \ \phi^4=\cos
f(r),\label{a2}\eea
where $r=|x|$ ($x\in\bfR^3$). Since the presence of the function $g$ gives rise to a nonvanishing temporal component of the gauge field,
$g$ may be regarded as an electric potential. With (\ref{a2}), the Skyrme charge $Q_S$ can be shown to be given by \cite{AT,BHT,LY}
\be \label{QS}
Q_S=-\frac2\pi\int_0^\infty \sin^2 f(r) f'(r)\,\dd r.
\ee

Recall also that, with the notation $\vec{\phi}=(\phi^\alpha)$ and the updated gauge-covariant derivative 
\be 
D_\mu\vec{\phi}=\pa_\mu{\vec\phi}+{\bf A}_\mu\times{\vec\phi},
\ee
we may express the 't Hooft electromagnetic field $F_{\mu\nu}$  by the formula \cite{JZ,tH0,tH} 
\be\label{EM}
F_{\mu\nu}=\frac1{|\vec{\phi}|}{\vec\phi}\cdot {\bf F}_{\mu\nu}-\frac1{|\vec{\phi}|^3}{\vec{\phi}}\cdot(D_\mu{\vec\phi}\times
D_\nu{\vec\phi}).
\ee

Inserting (\ref{a1}) and (\ref{a2}) into (\ref{EM}), we see that the electric and magnetic fields, 
${\bf E}=
(E^i)$ and ${\bf B}=(B^i)$, are given by \cite{Go,JZ,PS}
\bea
E^i&=&-F^{0i}=\frac{x^i}{r}\frac{\dd g}{\dd r},\label{g60}\\
B^i&=&-\frac12\epsilon_{ijk}F^{jk}=\frac{x^i}{r^3}.\label{g61}
\eea
 Therefore the magnetic  charge $Q_m$ may be calculated immediately to give us
\be\label{g62}
Q_m=\frac1{4\pi}\,\lim_{r\to\infty}\oint_{S^2_r}{\bf B}\cdot\dd{\bf S}=1,
\ee
where $S^2_r$ denotes the 2-sphere  of radius $r$, centered at the origin in the 3-space. Similarly,  the monopole charge $Q_M$ may be shown to be
1 as well \cite{LY}.

Within the ansatz (\ref{a1})--(\ref{a2}), using suitable rescaling, and denoting
$\kappa^4_2\equiv\kappa$, it is shown \cite{BHT} that the Lagrangian density (\ref{L}) may be reduced into the following one-dimensional one, after suppressing
the potential term,
\be \label{L1}
{\cal L}={\cal E}_1-{\cal E}_2,
\ee
where
\bea
{\cal E}_1&=&2\left(2(a')^2+\frac{(a^2-1)^2}{r^2}\right)+{1\over
2}\left(r^2(f')^2+2a^2\sin^2f\right)\nn\\
&&+2\kappa a^2\sin^2f\bigg(2f'^2+\frac{a^2\sin^2f}{r^2}\bigg),\label{E1}\\
{\cal E}_2&=&r^2(g')^2+2a^2g^2,\label{E2}\end{eqnarray}
and $'$ denotes the differentiation $\frac{\small\dd}{{\small\dd }r}$, such that the associated Hamiltonian (energy) density is given by
\be 
{\cal E}={\cal E}_1+{\cal E}_2.
\ee
The equations of motion of the original Lagrangian density (\ref{L}) now become the variational equation
\be \label{dA}
\delta {L}=0,
\ee
 of the static action 
\be \label{action}
{L}(a,f,g)=\int_0^\infty {\cal L}\,\dd r=\int_0^\infty({\cal E}_1-{\cal E}_2)\,\dd r,
\ee
 which is  indefinite. Explicitly, the equation (\ref{dA}) may be expressed in terms of
the unknowns $a,f,g$ as
\begin{eqnarray}
 a''&=&\frac{1}{r^2}a(a^2-1)+{1\over
 4}a\sin^2f+\kappa a\sin^2f(f')^2\nn\\
&&\quad +{1\over{r^2}}\kappa a^3\sin^4f
 -\frac{ag^2}{2},\label{4}\\
 8\kappa(a^2\sin^2ff')'+(r^2f')'&=&2a^2\sin f \cos f+8\kappa a^2\sin f\cos f(f')^2\nn\\
&&\quad+\frac{8\kappa a^4\sin^3f\cos f}{r^2},
\label{5}\\
(r^2g')'&=&2a^2g.\label{6}\end{eqnarray}

We are to solve these equations under suitable boundary conditions. First we observe in view of the ansatz (\ref{a1})--(\ref{a2}) that the regularity of the fields $\phi$ and $A_\mu$ 
imposes at $r=0$ the boundary condition
\be \label{bc1}
a(0)=1,\quad f(0)=\pi,\quad g(0)=0.
\ee 
Furthermore, the finite-energy condition
\be 
E(a,f,g)=\int_0^\infty{\cal E}\,\dd r=\int_0^\infty({\cal E}_1+{\cal E}_2)\,\dd r<\infty,
\ee
the definition of the vacuum manifold (\ref{V}), and the non-triviality of the $g$-sector lead us to the boundary condition at $r=\infty$, given as
\be \label{bc2}
a(\infty)=0,\quad f(\infty)=\omega,\quad g(\infty)=q,
\ee
where $q>0$ (say) is a parameter, to be specified later, which defined the asymptotic value of the electric potential at infinity.

Applying the boundary conditions (\ref{bc1}) and (\ref{bc2}) in (\ref{QS}), we obtain
\be \label{Qo}
Q_S=Q_S(\omega)=1+\frac1\pi \left(\frac12\sin(2\omega) -\omega\right),
\ee
which is strictly decreasing for $\omega\in [0,\pi]$ with $Q_S(0)=1, Q_S(\frac\pi2)=\frac12,Q_S(\pi)=0$, and the range of $Q_S(\omega)$ over $[0,\pi]$ is the entire interval $[0,1]$.

We now evaluate the electric charge.
Using (\ref{g60}) and the equation (\ref{6}), we see that the electric charge $Q_e$ is given by
\bea
Q_e&=&\frac1{4\pi}\lim_{r\to\infty}\oint_{S^2_r} {\bf E}\cdot\,\dd{\bf S}
=\frac1{4\pi}\,\lim_{r\to\infty}\int_{|x|<r}\nabla\cdot{\bf E}\,\dd x
=\frac1{4\pi}\,\lim_{r\to\infty}\int_{|x|<r}\pa_i\left(\frac{x^i}r\frac{\dd g}{\dd r}\right)\,\dd x\nn\\
&=&\int_0^\infty \frac{\dd}{\dd r}\left(r^2\frac{\dd g}{\dd r}\right)\,\dd r
=2\int_0^\infty a^2(r) g(r)\,\dd r.
\eea

With the above preparation, we can state our main result regarding the existence of dyon solitons in the minimally gauged Skyrme model \cite{BHT} as follows.

\begin{theorem} \label{Main} For any parameters $\omega$ and $q$ satisfying
\be \label{Co}
\frac\pi2<\omega<\pi,\quad 0<q<\min\left\{\frac1{\sqrt{2}}\sin\omega,\sqrt{2}\left(1-\frac\omega\pi\right)\right\},
\ee 
the equations of motion of the minimally gauged Skyrme model defined by the Lagrangian density (\ref{L}), with $\lm=0$, have a static
finite-energy spherically symmetric solution described by the ansatz (\ref{a1})--(\ref{a2}) so that $(a,f,g)$ satisfies the boundary
conditions (\ref{bc1}) and (\ref{bc2}), $a(r)>0, \omega<f(r)<\pi, 0<g(r)<q$ for all $r>0$, and $a,f,g$ are strictly monotone functions of $r$. Moreover,
$a(r)$ vanishes at infinity exponentially fast and $f(r), g(r)$ approach their limiting values at the rate {\rm O}$(r^{-1})$ as $r\to\infty$. The solution
carries a unit monopole charge, $Q_M=1$, a continuous Skyrme charge $Q_S$ given as function of $\omega$ by
\be 
Q_S(\omega)=1+\frac1\pi\left(\frac12\sin(2\omega)-\omega\right),\quad \frac\pi2<\omega<\pi,
\ee 
which may assume any value in the interval
$(0,\frac12)$, a unit magnetic charge $Q_m=1$, and an electric charge $Q_e$ given by the integral
\be \label{2.34}
Q_e=2\int_0^\infty a^2(r)g(r)\,{\rm\dd} r>0,
\ee
which depends on $q$ and approaches zero as $q\to0$.
\end{theorem}

It is interesting that the 't Hooft electric charge $Q_e$ cannot be quantized as stated in the Dirac quantization formula, which reads in normalized units \cite{Ryder},
\be \label{2.35}
q_e q_m =\frac n2,\quad n\in\bfZ,
\ee
where $q_e$ and $q_e$ are electric and magnetic charges, respectively. Indeed, according to
Theorem \ref{Main}, $Q_e=0$ is an accumulation point of the set of electric charges of the model. On the other hand,
the formula (\ref{2.35}) says that, for $q_m>0$, the smallest positive value of $q_e$ is $(2q_m)^{-1}$.
\medskip 

 We note that
the expression (\ref{2.34}) suggests that $Q_e$ should depend on $q$ continuously, although a proof of this statement is yet to be worked out.
\medskip 

The above theorem will be established in the subsequent sections.

\section{Constrained minimization problem}
\setcounter{equation}{0}

We first observe that the action density (\ref{L1}) is invariant under the transformation $f\mapsto\pi-f$. Hence we may `normalize' the boundary conditions (\ref{bc1}) and (\ref{bc2}) into
\bea 
a(0)&=&1,\quad f(0)=0,\quad g(0)=0,\label{bc3}\\
a(\infty)&=&0,\quad f(\infty)=\pi-\omega,\quad g(\infty)=q,\label{bc4}
\eea

The proof of our main existence theorem, Theorem \ref{Main}, for the dyon solutions in the minimally gauged Skyrme model amounts to establishing the following.

\begin{theorem}\label{theorem 01}
Given $\omega$ satisfying
\be \label{om}
\frac\pi2<\omega<\pi,
\ee
set
\be \label{qom}
q_\omega=\frac{\sqrt
2}{\pi}(\pi-\omega).
\ee
For any constant $q$ satisfying $0<q<q_\omega$ where $\omega$ lies in the interval (\ref{om}), and
\be \label{qsin}
q< \frac1{\sqrt{2}}\sin\omega,
\ee 
the functional
(\ref{action}) has a finite-energy critical point $(a,f,g)$ which satisfies the
equations (\ref{4})--(\ref{6}) and the boundary conditions
(\ref{bc3})--(\ref{bc4}). Furthermore, such a solution 
enjoys the property that $f(r),g(r)$ are strictly increasing, and
$a(r)>0, 0<f(r)<\pi-\omega, 0<g(r)<q$, for $r>0$. 
\end{theorem}

The proof of the theorem will be carried out through establishing a series of lemmas. In this section, we concentrate on formulating and solving
a constrained minimization problem put forth to overcome the difficulty arising from the negative terms in the action functional (\ref{action}).
In the next section, we show that the solution obtained in this section is indeed a critical point of (\ref{action}) so that the constraint does not
give rise to a Lagrangian multiplier problem.

To proceed, we begin by defining the admissible space of our one-dimensional variational problem to be
\bea
{\cal A}&=&\left\{(a,f,g)| a,f,g\mbox{ are continuous functions over $[0,\infty)$ which are
absolutely}\right.\nn\\
&&\,\left.\mbox{continuous on any compact subinterval of $(0,\infty)$,  satisfy}\right.\nn\\
&&\,\mbox{the boundary conditions $a(0)=1,a(\infty)=0,f(0)=0,f(\infty)=\pi-\omega,$}\nn\\
&&g(\infty)=q,
\left.\mbox{ and of finite-energy $E(a,f,g)<\infty$}\right\}.\nn
\eea

Note that in the admissible space $\cal A$ we only implement partially the boundary conditions (\ref{bc3})--(\ref{bc4}) to ensure the compatibility with the
minimization process. The full set of the boundary conditions will eventually be recovered in the solution process.

In order to tackle the problem arising from the negative terms involving the function $g$ in the action (\ref{action}), we use the methods developed in \cite{LY,SW,Yw} by imposing the
constraint
\begin{equation}\int_0^\infty(r^2g'G'+2a^2gG)\,\dd r=0, \label{8}\end{equation}
to `freeze' the troublesome $g$-sector, where $G$ is an arbitrary test function satisfying $G(\infty)=0$ and
\begin{equation}
E_2(a,G)=\int_0^\infty(r^2[G']^2+2a^2G^2)\,\dd r<\infty.\label{9}
\end{equation}
That is, for given $a$, the function $g$ is taken to be a critical point of the energy functional $E_2(a,\cdot)$ subject to the boundary condition $g(\infty)=q$.

We now define our constrained class $\mathcal{C}$ to be
\be \label{C}
{\cal C}=\{(a,f,g)\in {\cal A}|\, (a,f,g)  \mbox{ satisfies (\ref{8})}\}.
\ee

In the rest of this section, we shall study the following
constrained minimization problem
\begin{equation}
\min
\left\{L(a,f,g)|(a,f,g)\in
\mathcal{C}\right\}.\label{12}\end{equation}

\begin{lemma}\label{lemma0} Assume (\ref{om}). For the problem (\ref{12}), we may always restrict our attention to functions $f$ satisfying $0\leq f\leq \pi/2$.
\end{lemma}

\begin{proof} Since the action (\ref{action}) is even in $f$, it is clear that $L(a,f,g)=L(a,|f|,g)$. Hence we may assume $f\geq0$ in the minimization problem. Besides, since
$f(\infty)=\pi-\omega<\frac\pi2$, if there is some $r_0>0$ such that $f(r_0)>\frac\pi2$, then there is an interval $(r_1,r_2)$ with $0\leq r_1<r_0<r_2<\infty$ such that $f(r)>\frac\pi2$ ($r\in (r_1,r_2)$) and $f(r_1)=f(r_2)=\frac\pi2$. We now modify $f$ by reflecting $f$ over the interval $[r_1,r_2]$ with respect to the level $\frac\pi2$ to get a new function $\tilde{f}$ satisfying
$\tilde{f}(r)=\pi-f(r)$ ($r\in [r_1,r_2]$) and $\tilde{f}(r)=f(r)$ ($r\not\in[r_1,r_2]$). We have $L(a,f,g)\geq L(a,\tilde{f},g)$ again.
\end{proof}

\begin{lemma}{\label{lemma 1}}
The constrained admissible class $\cal C$ defined in (\ref{C}) is non-empty. Furthermore, if $q>0$ and $(a,f,g)\in{\cal C}$, we have $0<g(r)<q$ for all $r>0$ and that $g$ is the unique solution
to the minimization problem
\be \min \left\{E_2(a,G)\, \bigg|\, G(\infty)=q
\right\}.\label{01}\ee
 \end{lemma}
 \begin{proof}
Consider the problem (\ref{01}).
 Then the Schwartz inequality gives us the asymptotic estimate
\be
|G(r)-q|\leq \int_r^\infty\left|G'(\rho)\right|\,\dd\rho
\leq r^{-\frac12}\left(\int_r^\infty
\rho^2(G'(\rho))^2\,\dd\rho\right)^{1\over 2}
\leq r^{-{1\over 2}}E_2^{\frac12}(a,G),
\label{10}\ee 
which indicates that the limiting behavior $G(\infty)=q$ can be preserved for any minimizing sequence of the problem (\ref{01}). Hence (\ref{01}) is solvable. In fact, it has
a unique solution, say $g$, for any given function $a$, since the functional $E_2(a,\cdot)$ is strictly convex. Since $E_2(a,\cdot)$ is even, we have $g\geq0$.
Applying the maximum principle in (\ref{6}), we conclude with $0<g(r)<q$ for all $r>0$. The uniqueness of the solution to (\ref{01}), for given $a$, is obvious.
\end{proof}

\begin{lemma}{\label{lemma 2}} For any $(a,f,g)\in
\mathcal{C}$, $g(r)$ is nondecreasing for $r>0$ and
$g(0)=0.$\end{lemma}
\begin{proof} 
To proceed, we first claim that 
\be
\liminf\limits_{r\rightarrow 0}r^2|g'(r)|=0.\label{02}
\ee
Indeed, if (\ref{02}) is false, then there are
$\epsilon_0,\delta>0$, such that $r^2|g'(r)|\geq \epsilon_0$ for
$0<r<\delta$, which contradicts the convergence of the integral
$\int_0^\infty r^2(g')^2 \dd r$.

As an immediate consequence, (\ref{02}) implies that there is a sequence $\{r_k\}$, such that
$r_k\rightarrow 0$ and $r_k^2|g'(r_k)|\rightarrow 0$, as
$k\rightarrow \infty$.
In view of this fact and (\ref{6}), we
have
\bea
r^2g'(r)&=&r^2g'(r)-\lim\limits_{k\rightarrow
\infty}r_k^2g'(r_k)\nn\\
&=&\lim\limits_{k\rightarrow
\infty}\int_{r_k}^r(\rho^2g'(\rho))'\,\dd\rho=\int_0^r(\rho^2g'(\rho))'\,\dd\rho\nn\\
&=&\int_0^r2a^2(\rho)g(\rho)\,\dd\rho\geq 0,\quad r>0.\label{3.12}
\eea
Hence $g'(r)\geq 0$ and
$g(r)$ is nondecreasing.
In particular, we conclude that there is number $g_0\geq0$ such that 
\be\lim\limits_{r\rightarrow
0}g(r)=g_0.\label{04}\ee
We will need to show $g_0=0$.
Otherwise, if $g_0>0$, we can use $a(0)=1,r^2g'(r)\rightarrow
0$ ($r\rightarrow 0$) (this latter result follows from (\ref{3.12})), and L'Hopital's rule to get
$$2g_0=2\lim\limits_{r\rightarrow
0}a^2(r)g(r)=\lim\limits_{r\rightarrow
0}(r^2g')'=\lim\limits_{r\rightarrow
0}\frac{r^2g'(r)}{r}=\lim\limits_{r\rightarrow 0}rg'(r).$$ Hence,
there is a $\delta>0$, such that \be g'(r)\geq \frac{g_0}r,\quad
0<r<\delta.\label{03}\ee
Integrating (\ref{03}), we obtain $$\left
|g(r_2)-g(r_1)\right|\geq g_0\left|\ln \frac{r_2}{r_1}\right|,$$
which contradicts the existence of limit stated in (\ref{04}). So $g_0=0,$
and the lemma follows.\end{proof}

\begin{lemma}{\label{lemma 3}} With (\ref{om}) and (\ref{qom}), for
any $0<q<q_\omega$,
\be \label{qq}
q<\frac1{\sqrt{2}}(\pi-\omega),
\ee 
 and $(a,f,g)\in {\cal C}$, we have the following partial coercive lower estimate
\bea 
L(a,f,g)&\geq&
\int_0^\infty
{\rm\dd} r\left\{2\left(2(a')^2+\frac{(a^2-1)^2}{r^2}\right)+C_1 r^2(f')^2\right.\nn\\
&&\left.+2\kappa
a^2\sin^2f\bigg(2(f')^2
+\frac{a^2\sin^2f}{r^2}\bigg)+C_2 a^2f^2\right\},\label{11}
\eea
where $C_1,C_2>0$ are constants depending on $\omega$ and $q$ only.
\end{lemma}
\begin{proof}
 For any $(a,f,g)\in
\mathcal{C}$, set $g_1={q}{(\pi-\omega)^{-1}}f$. Then $g_1$ satisfies
$g_1(\infty)=q$. As a consequence, we have
\be 
E_2(a,g_1)\geq E_2(a,g),
\ee
and thus,
\begin{eqnarray}\label{3.17}
&&L(a,f,g)=E_1(a,f)-E_2(a,g)
\geq E_1(a,f)-E_2(a,g_1)\nonumber\\
&&=\int_0^\infty
\dd r\left\{2\left(2(a')^2+\frac{(a^2-1)^2}{r^2}\right)+\left({1\over
2}-\frac{q^2}{(\pi-\omega)^2}\right)r^2(f')^2\right.\nonumber\\
&&+2\kappa
a^2\sin^2f\bigg(2(f')^2+\frac{a^2\sin^2f}{r^2}\bigg)\left.+\left(\frac{\sin^2f}{f^2}-\frac{2q^2}{(\pi-\omega)^2}\right)a^2f^2\right\}.\end{eqnarray}

Using the elementary inequality $\frac{\sin t}t\geq\frac2\pi$ ($0<t\leq \frac\pi2$) and Lemma \ref{lemma0}, we have
\bea 
\left(\frac{\sin^2f}{f^2}-\frac{2q^2}{(\pi-\omega)^2}\right)f^2&\geq& 2\left(\frac2{\pi^2}-\frac{q^2}{(\pi-\omega)^2}\right)f^2\nn\\
&=&\frac2{(\pi-\omega)^2}(q_\omega^2-q^2)f^2\equiv C_2 f^2.\label{3.18}
\eea
Inserting (\ref{3.18}) into (\ref{3.17}) and setting in (\ref{3.17}) the quantity
\be 
C_1\equiv {1\over
2}-\frac{q^2}{(\pi-\omega)^2}=\frac1{(\pi-\omega)^2}\left(\frac12 (\pi-\omega)^2-q^2\right)>0,
\ee 
in view of (\ref{qq}), we see that the lower estimate (\ref{11}) is established.
\end{proof}

\begin{lemma}{\label{lemma 4}}
Under the conditions stated in Theorem \ref{theorem 01}, the constrained minimization problem (\ref{12}) has a solution.
\end{lemma}
\begin{proof} We start by observing that the condition (\ref{qq}) is implied by the condition (\ref{qsin}). So Lemma \ref{lemma 3} is valid.
Hence, applying Lemma \ref{lemma 3}, we see that
\be
\eta=\inf\{L(a,f,g)\,|\,(a,f,g)\in {\cal C}\}
\ee
is well defined.
Let $\{(a_n,f_n,g_n)\}$ denote any minimizing
sequence of (\ref{12}). That is, $(a_n,f_n,g_n)\in {\cal C}$ and $L(a_n,f_n,g_n)\to\eta$ as $n\to\infty$. Without loss of generality, we may assume
$L(a_n,f_n,g_n)\leq \eta+1$ (say) for all $n$.
In view of (\ref{11}) and the Schwartz inequality, we have
\be\label{3.21}
|a_n(r)-1|\leq\int_0^r\left|a'_n(\rho)\right|\,\dd \rho\leq r^{\frac12}\left(\int_0^r (a_n'(\rho))^2\,\dd\rho\right)^{\frac12}\leq C r^{\frac12}(\eta+1)^{\frac12},
\ee
\be\label{3.22}
\left|f_n(r)-(\pi-\omega)\right|\leq\int_r^\infty\left|f'_n(\rho)\right|\,\dd\rho
\leq r^{-\frac12}\left(\int_r^\infty \rho^2 (f'_n(\rho))^2\,\dd\rho\right)^{\frac12}\leq Cr^{-\frac12}(\eta+1)^{\frac12},
\ee
where $C>0$ is a constant independent of $n$.
In particular,
$a_n(r)\rightarrow 1$ and $f_n(r)\rightarrow (\pi-\omega)$
uniformly as $r\rightarrow 0$ and $r\rightarrow \infty$,
respectively.

For any $(a_n,f_n,g_n)$, the function
$G_n=\frac{{q}}{{(\pi-\omega)}}f_n$ satisfies $G_n(\infty)=q$.
Thus, by virtue of the definition of $g_n$ and (\ref{11}), we have
 \be \label{3.23}
E_2(a_n,g_n)\leq
E_2(a_n,G_n)
=\frac{q^2}{(\pi-\omega)^2}\int_0^\infty(r^2(f'_n)^2+a_n^2f_n^2)\,\dd r\leq CL(a_n,f_n,g_n),
\ee
where $C>0$ is a constant, which shows that
$E_2(a_n,f_n)$ is bounded as well.

With the above preparation, we are now ready to investigate the limit of the sequence $\{(a_n,f_n,g_n)\}$.

Consider the Hilbert space $(X,(\cdot,\cdot))$, where the
 functions in $X$ are all continuously defined in $r\geq0$ and
 vanish at $r=0$ and the inner product $(\cdot,\cdot)$ is defined
 by $$(h_1,h_2)=\int_0^\infty h'_1(r)h'_2(r)\,\dd r, \ \ h_1,h_2\in X.$$

Since $\{a_n-1\}$ is bounded in $(X,(\cdot,\cdot))$, we may assume
without loss of generality that $\{a_n\}$ has a weak limit, say,
$a$, in the same space,
\begin{equation}
\int_0^\infty a'_n h'\,\dd r\rightarrow\int_0^\infty
a' h'\,\dd r,\ \ \ \forall h\in
X,\nonumber\end{equation} as $n\rightarrow \infty$.

Similarly, for the Hilbert space $(Y,(\cdot,\cdot))$ where the
functions in  $Y$ are all continuously defined in $r>0$ and vanish
at infinity and the inner product $(\cdot,\cdot)$ is defined by
$$(h_1,h_2)=\int_0^\infty r^2h_1'h_2'\,\dd r, \ \ h_1,h_2\in Y.$$
Since $\{f_n-(\pi-\omega)\}$, $\{g_n-q\}$ are bounded in
$(Y,(\cdot,\cdot))$, we may assume without loss of generality that there are functions $f,g$ with
$f(\infty)=\pi-\omega, g(\infty)=q$, and $f-(\pi-\omega), g-q\in
(Y,(\cdot,\cdot))$, such that
\begin{equation}
\int_0^\infty r^2H_n'h'\,\dd r\rightarrow
\int_0^\infty r^2H'h'\,\dd r,\quad
\forall h\in Y,\label{15}\end{equation}
as $n\to\infty$, for
$H_n=f_n-(\pi-\omega),\ H=f-(\pi-\omega)$, and $H_n=g_n-q,\ H=g-q$, respectively.

Next, we need to show that the weak limit $(a,f,g)$ of the minimizing
sequence $\{(a_n,f_n,g_n)\}$ obtained above actually lies in
$\mathcal{C}$. There are two things to be verified for $(a,f,g)$: the boundary conditions and the constraint (\ref{8}).
From the uniform estimates (\ref{10}), (\ref{3.21}), and (\ref{3.22}), we easily deduce that $a(0)=1,f(\infty)=\pi-\omega,g(\infty)=q$.
Moreover,
applying Lemma \ref{lemma 3}, we get $a\in W^{1,2}(0,\infty)$.  Hence $a(\infty)=0$. To verify $f(0)=0$, we use (\ref{3.21}) to get a $\delta>0$ such that 
\be \label{an}
|a_n(r)|\geq\frac12,\quad r\in[0,\delta].
\ee
Then, using (\ref{an}), we have
\bea \label{fn}
\sin^2 f_n(r)&\leq&2\int_0^r|\sin f_n(\rho) f'_n(\rho)|\,\dd\rho\nn\\
&\leq&4 r^{\frac12}\left(\int_0^r a_n^2(\rho)\sin^2 f_n(\rho) (f'_n(\rho))^2\,\dd\rho\right)^{\frac12}\nn\\
&\leq& 2\kappa^{-\frac12}r^{\frac12} L^{\frac12}(a_n,f_n,g_n),\quad r\in[0,\delta].
\eea
Since $0\leq f_n\leq\frac\pi2$, we can invert (\ref{fn}) to obtain the uniform estimate
\be \label{fnn}
0\leq f_n(r)\leq Cr^{\frac14},\quad r\in[0,\delta],
\ee
where $C>0$ is independent of $n$. Letting $n\to\infty$ in (\ref{fnn}), we see that $f(0)=0$ as anticipated.

Thus, it remains to verify (\ref{8}). For this purpose, it suffices to establish the following
results,
\bea 
\int_0^\infty(a^2_ng_n-a^2g)G\,
\dd r &\rightarrow& 0,\label{*}\\
\ \int_0^\infty
(r^2g'_n-r^2g')G' \,\dd r&\rightarrow& 0,\label{**}
\eea
for any test function $G$ satisfying (\ref{9}) and $G(\infty)=0$, as $n\to\infty$.

From the fact $G\in Y$ and (\ref{15}), we immediately see that (\ref{**}) is valid.

To establish (\ref{*}), we
rewrite 
\be 
 \int_0^\infty(a^2_ng_n-a^2g)G\,
\dd r=\int_0^{\delta_1}+\int_{\delta_1}^{\delta_2}+\int_{\delta_2}^\infty
\equiv I_1+I_2+I_3,
\ee
for some positive constants $0<\delta_1<\delta_2<\infty$, and we begin with
\be 
I_1
=\int_0^{\delta_1}(a_n^2-a^2)g_n G\,
\dd r+\int_0^{\delta_1}a^2(g_n-g)G\, \dd r
\equiv I_{11}+I_{12}.
\ee
In view of (\ref{3.21}) and (\ref{3.23}), we see that there is a small $\delta>0$ such that $g_n\in L^2(0,\delta)$ and there holds the uniform bound
\be 
\|g_n\|_{L^2(0,\delta)}\leq K,
\ee
for some constant $K>0$. Thus, we may assume $g_n\to g$ weakly in $L^2(0,\delta)$ as $n\to\infty$. In particular, $g\in L^2(0,\delta)$ and $\|g\|_{L^2(0,\delta)}\leq K$.
Besides, since in (\ref{9}), the function $a$ satisfies $a(0)=1$, we have $G\in L^2(0,\delta)$ when $\delta>0$ is chosen small enough. Thus, using (\ref{3.21}) and taking $\delta_1\leq\delta$, we get
\begin{eqnarray}
|I_{11}|&\leq&\int_0^{\delta_1}|a^2_n-a^2|
|g_n G|\,\dd r
\leq\int_0^{\delta_1}\left(|a^2_n-1|+|a^2-1|\right)|g_n G|\,\dd r\nonumber\\
&\leq& CK \delta^{\frac12}\|G\|_{L^2(0,\delta)},
\end{eqnarray}
where $C>0$ is a constant independent of $n$. 
Thus, for any $\vep>0$, we can choose $\delta_1>0$ sufficiently small to get $|I_{11}|<\vep$.
On the other hand,
since $g_n\to g$ weakly in $L^2(0,\delta)$ and $G\in L^2(0,\delta)$, we have $I_{12}\to0$ as $n\to\infty$.

Since $\{a_n\}$ and $\{g_n\}$ are bounded sequences in $W^{1,2}(\delta_1,\delta_2)$, using the compact embedding
$W^{1,2}(\delta_1,\delta_2)\mapsto C[\delta_1,\delta_2]$, we see that $a_n\to a$ and $ g_n\to g$ uniformly over $[\delta_1,\delta_2]$ as $n\to\infty$.
Thus $I_2\to0$ as $n\to\infty$.

To estimate $I_3$, we recall that $\{E_2(a_n,g_n)\}$ is bounded by (\ref{3.23}), 
$g_n(r)\rightarrow q$ uniformly as $n\to\infty$ by (\ref{10}), and $G(r)=\mbox{O}(r^{-\frac12})$ as
$r\to \infty$ by (\ref{9}). In particular, since $q>0$, we may choose $r_0>0$ sufficiently large so that
\be \label{3.32}
|g(r)|\geq\frac q2,\quad \inf_{n}|g_n(r)|\geq \frac q2,\quad r\geq r_0.
\ee
Combining the above facts, we arrive at
\be \label{3.33}
|I_3|\leq\int_r^\infty \left(|a^2_n g_n |+|a^2 g|\right)|G|\,\dd \rho\leq Cr^{-\frac12}\int^\infty_r\frac2q (a^2_n g_n^2+a^2 g^2)\,\dd\rho,
\ee
where $r\geq r_0$ (cf. (\ref{3.32})) and $C>0$ is a constant. Using (\ref{3.23}) in (\ref{3.33}), we see that for any $\vep>0$ we may choose $\delta_2$ large enough
to get $|I_3|<\vep$.

Summarizing the above discussion, we obtain
\be 
\limsup_{n\to\infty}\left|\int_0^\infty(a^2_ng_n-a^2g)G\,
\dd r\right|\leq 2\vep,
\ee
which proves the desired conclusion (\ref{*}). Thus, the claim $(a,f,g)\in{\cal C}$ follows.

To show that $(a,f,g)$ solves (\ref{12}), we need to establish
\be
\eta=\liminf_{n\to\infty} L(a_n,f_n,g_n)\geq L(a,f,g).
\label{00}\ee
This fact is not automatically valid and extra caution is to be exerted because the functional $L$ contains negative terms.

With
\be 
\Om=\pi-\omega,\quad
0<\Om<\frac\pi2,
\ee 
we may rewrite the Lagrange density (\ref{L1}) as
\be
{\mathcal{L}}(a,f,g)
= {\cal L}_0(a,f)-{\cal E}_0(a,g),\ee
where
\bea 
{\cal L}_0(a,f)&=&2\bigg(2(a')^2+\frac{(a^2-1)^2}{r^2}\bigg)+{1\over
2}r^2(f')^2+2\kappa a^2\sin^2f\bigg(2(f')^2+\frac{a^2\sin^2f}{r^2}\bigg)\nonumber\\
&&+a^2(\sin^2\Om-2q^2)+a^2\sin^2\Om\left(\cos^2(f-\Om)-1\right)\nonumber\\
&&+2a^2\sin\Om\cos\Om\sin(f-\Om)\cos(f-\Om)
+a^2\cos^2\Om\sin^2(f-\Om),\\
{\cal E}_0(a,g)&=&r^2(g')^2+2a^2(g-q)^2+4a^2(g-q)q,
\eea 
 Thus, in order to establish (\ref{00}), it suffices to show that
\be \label{L0}
\liminf_{n\to\infty}\int_0^\infty {\cal L}_0(a_n,f_n)\,\dd r\geq \int_0^\infty{\cal L}_0(a,f)\,\dd r,
\ee
\be \label{E0}
\lim_{n\to \infty}\int_0^\infty {\cal E}_0(a_n,g_n)\,\dd r=\int_0^\infty{\cal E}_0(a,f)\,\dd r.
\ee

We first show (\ref{E0}). To this end, we observe that, 
since both $(a_n,g_n)$ and $(a,g)$ satisfy (\ref{8}), i.e.,
\be
\int_0^\infty(r^2g'_n G'+2a^2_ng_n G)\,\dd r =0,\quad
\int_0^\infty(r^2g' G'+2a^2g G)\,\dd r=0,
\ee
we can set $G=g-g_n$ in the above equations and
subtract them to get
\bea\label{3I}
 \int_0^\infty
r^2(g'_n-g')^2\dd r&=&2\int_0^\infty(a^2_ng_n-a^2g)(g-g_n)\dd r\nn\\
&=&\int_0^{\delta_1}+\int_{\delta_1}^{\delta_2}+\int_{\delta_2}^\infty\equiv I_1+I_2+I_3,
\eea
where $0<\delta_1<\delta_2<\infty$.

To study $I_1$, we need to get some uniform estimate for the sequence $\{g_n\}$ near $r=0$. From (\ref{3.21}), we see that for any $0<\gamma<\frac12$ (say) there is a $\delta>0$ such that
\be 
2a^2_n(r)\geq (2-\gamma),\quad r\in[0,\delta].
\ee
Consider the comparison function
\be \label{sig}
\sigma(r)=C r^{1-\gamma},\quad r\in[0,\delta], \quad C>0.
\ee
Then
\be 
(r^2 \sigma')'=(1-\gamma)(2-\gamma)\sigma<2a^2_n(r)\sigma,\quad r\in[0,\delta].
\ee
Consequently, we have
\be \label{3.47}
(r^2(g_n-\sigma)')'>2a^2_n(r)(g_n-\sigma),\quad r\in[0,\delta].
\ee 
Choose $C>0$ in (\ref{sig}) large enough so that $C\delta^{1-\gamma}\geq q$. Since $g_n<q$ (Lemma \ref{lemma 1}), we have $(g_n-\sigma)(\delta)<0$ and $(g_n-\sigma)(0)=0$. In view of these
boundary conditions and applying the maximum principle to (\ref{3.47}), we obtain $g_n(r)<\sigma(r)$ for all $r\in(0,\delta)$. Or, more precisely, we have
\be \label{Eg}
0<g_n(r)<\left(\frac q{\delta^{1-\gamma}} \right) r^{1-\gamma},\quad 0<r<\delta.
\ee 
Of course, the weak limit $g$ of $\{g_n\}$ satisfies the same estimate. Therefore, using the uniform estimates (\ref{3.21}) and (\ref{Eg}), we see that for any $\vep>0$ there is some $\delta_1>0$ ($\delta_1<\delta$)
such that $|I_1|<\vep$.

Moreover, in view of the uniform estimate (\ref{10}) and (\ref{3.32}), we have
\bea 
|I_3|&\leq&2\int_{\delta_2}^\infty (a_n^2g_n+a^2 g)(|g_n-q|+|g-q|)\,\dd r\nn\\
&\leq&\frac4q(|g_n(\delta_2)-q|+|g(\delta_2)-q|)\int_0^\infty (a_n^2 g_n^2+a^2 g^2)\,\dd r\nn\\
&\leq&\frac2q \delta_2^{-\frac12} \left(E_2^{\frac12}(a_n,g_n)+E_2^{\frac12}(a,g)\right)\left(E_2(a_n,g_n)+E_2(a,g)\right),
\eea
which may be made small than $\vep$ when $\delta_2>0$ is large enough due to (\ref{3.23}).

Furthermore, since $a_n\to a$ and $g_n\to g$ in $C[\delta_1,\delta_2]$, we see that $I_2\to0$ as $n\to\infty$.

In view of the above results regarding $I_1, I_2, I_3$ in (\ref{3I}), we obtain the strong convergence
\begin{equation}\lim\limits_{n\rightarrow\infty}\int_0^\infty
r^2(g'_n-g')^2\,\dd r=0.\end{equation} In particular, we have
\be \label{3.51}
\lim_{n\to\infty}\int_0^\infty r^2 (g'_n)^2\,\dd r=\int_0^\infty r^2 (g')^2\,\dd r.
\ee

We can also show that
\be \label{3.52}
\lim_{n\to\infty}\int_0^\infty\left(a_n^2 (g_n-q)^2+2a_n^2(g_n-q)q\right)\,\dd r=\int_0^\infty\left(a^2 (g-q)^2+2a^2(g-q)q\right)\,\dd r.
\ee 

In fact, we have seen that $\{(a_n,f_n,g_n)\}$ is bounded in $W^{1,2}_{\mbox{loc}}(0,\infty)$. Thus, the sequence is convergent in $C[\alpha,\beta]$ for any
pair of numbers, $0<\alpha<\beta<\infty$. Since we have shown that $a_n(r)\to 1$ and $g_n(r)\to 0$ as $r\to0$ uniformly, with respect to $n=1,2,\cdots$, we conclude that
$a_n\to a$ and $g_n\to g$ uniformly over any interval $[0,\beta]$ ($0<\beta<\infty$). Thus, combining this result with the uniform estimate (\ref{10}), we see that
(\ref{3.52}) is proved.

In view of (\ref{3.51}) and (\ref{3.52}), we see that (\ref{E0}) follows.

On the other hand, applying the uniform estimate (\ref{3.22}), we also have
\bea 
\lim_{n\to\infty}\int_0^\infty a_n^2 (\cos^2(f_n-\Om)-1)\,\dd r &=& \int_0^\infty a^2 (\cos^2(f-\Om)-1)\,\dd r,\label{lim1}\\
\lim_{n\to\infty}\int_0^\infty a_n^2\sin(f_n-\Om)\cos(f_n-\Om)\,\dd r&=&\int_0^\infty a^2\sin(f-\Om)\cos(f-\Om)\,\dd r.\nn\\ \label{lim2}
\eea

Finally, using (\ref{lim1}), (\ref{lim2}), and the condition (\ref{qsin}), i.e.,
\be 
\sin^2\Om-2q^2> 0,
\ee
we see that (\ref{L0}) is established and the proof of the lemma is complete.
\end{proof}

\section{Fulfillment of the governing equations}
\setcounter{equation}{0}

Let $(a,f,g)$ be the solution of (\ref{12}) obtained in the previous section. We need to show that it satisfies the governing equations (\ref{4})--(\ref{6}) for dyons.
Since we have solved a constrained minimization problem, we need to prove that the Lagrange multiplier problem does not arise
as a result of the constraint, which would otherwise alter the original
equations of motion. In fact, since the constraint (\ref{8}) involves $a$ and $g$ only and (\ref{8}) immediately gives rise to (\ref{6}), we see that all we have to do is
to verify the validity of (\ref{4}) because (\ref{5}) is the $f$-equation and (\ref{8}) does not involve $f$ explicitly.

To proceed, we take $\widetilde{a}\in C^1_0$. For any $t\in \bfR$,
there is a unique corresponding function $g_t$ such that
$(a+t\widetilde{a},f,g_t)\in \mathcal{C}$ and that $g_t$ smoothly
depends on $t$. Set
\be
g_t=g+\widetilde{g}_t, \quad \widetilde{G}=\left.\left(\frac{\dd}{\dd t}\widetilde{g}_t\right)\right|_{t=0}.
\ee
Since $(a+t\widetilde{a},f,g_t)|_{t=0}=(a,f,g)$ is a minimizing
solution of  (\ref{12}), 
we have
\bea \label{4.2}
0&=&\left.\frac{\dd}{\dd t}L(a+t\widetilde{a},f,g_t)\right|_{t=0}\nn\\
&=&8\int_0^\infty \dd r\bigg\{a'\widetilde{a}'+\frac{(a^2-1)a\widetilde{a}}{r^2}+\frac14\sin^2f \, a\widetilde{a}\nn\\
&&+\kappa\sin^2f\bigg((f')^2a\widetilde{a}+\frac{\sin^2f}{r^2}a^3\widetilde{a}\bigg)
-\frac12g^2a\widetilde{a}\bigg\}
-2\int_0^\infty\dd r\bigg\{r^2g'\widetilde{G}'+2a^2g\widetilde{G}\bigg\}\nn\\
&\equiv& 8I_1-2I_2.
\eea

It is clear that the vanishing of $I_1$ implies (\ref{4}) so that it suffices to prove that $I_2$ vanishes. To this end and in view of (\ref{8}), we only need to show that $\widetilde{G}$
satisfies the same conditions required of $G$ in (\ref{8}).

In (\ref{8}), when we make the replacements $a\mapsto a+t\widetilde{a},g\mapsto
g_t,G\mapsto \widetilde{g}_t$, we have
\begin{equation}\int_0^\infty\left(r^2g'_t\widetilde{g}'_t+2(a+t\widetilde{a})^2g_t\widetilde{g}_t\right)\,\dd r=0.\end{equation}
Or, with $g_t=g+\widetilde{g}_t$, we have
\begin{equation}\label{4.4}
\int_0^\infty
\left(r^2(g'+\widetilde{g}'_t)\widetilde{g}'_t+2a^2(g+\widetilde{g}_t)\widetilde{g}_t+2t^2\widetilde{a}^2g_t\widetilde{g}_t+4ta\widetilde{a}g_t\widetilde{g}_t\right)\,\dd r=0.
\end{equation}  
Recall that $\int_0^\infty\left(
r^2g'\widetilde{g}'_t+2a^2g\widetilde{g}_t\right)\,\dd r=0$. Thus (\ref{4.4}) and the Schwartz inequality give us
\begin{eqnarray}
&&\int_0^\infty(r^2(\widetilde{g}'_t)^2+2a^2\widetilde{g}^2_t)\,\dd r
=\left|2t\int_0^\infty(t\widetilde{a}^2+2a\widetilde{a})g_t\widetilde{g}_t\,\dd r\right|\nonumber\\
&&\leq|2t|\left(|2t|\int_0^\infty \widetilde{a}^2g_t^2
\dd r+\frac{1}{|2t|}\int_0^\infty
a^2\widetilde{g}^2_t\,\dd r\right)+2t^2\int_0^\infty
\widetilde{a}^2|g_t|\,|\widetilde{g}_t|\,\dd r\nonumber\\
&=&4t^2\int_0^\infty \widetilde{a}^2g^2_t \,\dd r+\int_0^\infty
a^2\widetilde{g}^2_t\,\dd r+2t^2\int_0^\infty
\widetilde{a}^2|g_t|\,|\widetilde{g}_t|\,\dd r.\label{4.5}
\end{eqnarray}
Applying the bounds $0\leq g, g_t\leq q$ and the relation $\widetilde{g}_t=g_t-g$ in (\ref{4.5}), we have
 \begin{eqnarray}
\int_0^\infty
(r^2(\widetilde{g}'_t)^2+a^2\widetilde{g}^2_t)\,\dd r&\leq&4t^2\int_0^\infty
\widetilde{a}^2g^2_t\,\dd r+2t^2\int_0^\infty
\widetilde{a}^2|g_t|\,|\widetilde{g}_t|\,\dd r \nonumber\\
&\leq& 8q^2 t^2\int_0^\infty
\widetilde{a}^2\,\dd r.\label{4.6}
\end{eqnarray}
As a consequence,
we have 
 \be
\int_0^\infty\left(r^2\left(\frac{\widetilde{g}'_t}{t}\right)^2+a^2\left(\frac{\widetilde{g}_t}{t}\right)^2\right)\,\dd r
\leq 8q^2\int_0^\infty \widetilde{a}^2\,\dd r,\quad t\neq 0.\label{17}\ee
Using
$\widetilde{g}_t(\infty)=0$, the Schwartz inequality and (\ref{17}), we have for $t\neq0$ the estimate
\be\label{4.8}
\left|\frac{\widetilde{g}_t}{t}(r)\right|\leq \int_r^\infty\left|
\frac{\widetilde{g}_t'(\rho)}{t}\right|\,\dd\rho\nonumber\\
\leq r^{-\frac{1}{2}}\left(\int_r^\infty
\rho^2\left(\frac{\widetilde{g}'_t}{t}\right)^2\,\dd\rho\right)^{\frac{1}{2}}\leq 2\sqrt{2} q\|\widetilde{a}\|_{L^2(0,\infty)}.
\ee

Letting $t\to0$ in (\ref{17}) and (\ref{4.8}), we obtain $E_2(a,\widetilde{G})<\infty$ and $\widetilde{G}(r)=\mbox{O}(r^{-\frac12})$ (for $r$ large). In particular,
$\widetilde{G}(\infty)=0$ and $\widetilde{G}$ indeed satisfies all conditions required in (\ref{8}) for $G$. Hence $I_2$ vanishes in (\ref{4.2}).
Consequently, the equation (\ref{4}) has been verified.

\section{Properties of the solution obtained}
\setcounter{equation}{0}

In this section, we study the properties of the solution, say $(a,f,g)$, of the equations (\ref{4})--(\ref{6}) obtained as a solution of the constrained minimization
problem (\ref{12}). We split the investigation over a few steps.

\begin{lemma}{\label{lemma 6}}
The solution $(a,f,g)$ enjoys the properties $a(r)>0$, $0<g(r)<q$, $0<f(r)<\pi-\omega$, and both $f(r)$ and $g(r)$ are strictly increasing, for any $r>0$.
\end{lemma}
\begin{proof}
We have $0\leq g\leq q$ and $0\leq f\leq \frac\pi2$ from Lemmas \ref{lemma0} and \ref{lemma 1}. Besides, it is clear that $a\geq0$ since
both (\ref{E1}) and (\ref{E2}) are even in $a$.

If $a(r_0)=0$ for some $r_0>0$, then $r_0$ is a
minimizing point and $a'(r_0)=0$. Using the uniqueness of the solution to the initial value problem consisting of
(\ref{4}) and $a(r_0)=a'(r_0)=0$, we get $a\equiv0$ which contradicts $a(0)=1$. Thus, $a(r)>0$ for all $r>0$.
The same argument shows that $f(r)>0, g(r)>0$ for all $r>0$. Since (\ref{3.12}) is valid, we see that $g(r)$ is strictly increasing. In particular, $g(r)<q$ for all $r>0$.

Lemma \ref{lemma0} already gives us $f\leq\frac\pi2$. We now strengthen it to $f<\pi-\omega$. First it is easy to see that $f\leq\pi-\omega$. Otherwise there is
a point $r_0>0$ such that $f(r_0)>\pi-\omega$. Thus, we can find two points $r_1, r_2$, with $0\leq r_1<r_0<r_2$, such that $f(r_1)=f(r_2)$ and $f(r)\geq f(r_1)$
for all $r\in (r_1,r_2)$. Modify $f$ to $\tilde{f}$ by setting $\tilde{f}(r)=f(r_1)$, $r\in (r_1,r_2)$; $\tilde{f}=f$, elsewhere. Then $(a,\tilde{f},g)\in {\cal C}$ and
$L(a,\tilde{f},g)<L(a,f,g)$ because $f$ cannot be constant-valued over $(r_1,r_2)$ by virtue of the equation (\ref{5}) and the energy density ${\cal E}_1$ defined in
(\ref{E1}) increases for $f\in [0,\frac\pi2]$. This contradiction establishes the result $f\leq\pi-\omega$.
Next, we assert that $f<\pi-\omega$. Otherwise, if
$f(r_0)=\pi-\omega$ for some $r_0>0$, then $r_0$ is a 
maximum point of $f$ such that $f'(r_0)=0$ and $f''(r_0)\leq 0$. Inserting these results
into (\ref{5}), we arrive at a contradiction since $0<\pi-\omega<\frac\pi2$. 

To see that $f$ is non-decreasing, we assume otherwise that there are $0<r_1<r_2$ such that $f(r_1)>f(r_2)$. Since $f(0)=0$, we see that $f$ has a local maximum point $r_0$ below
$r_2$, which is known to be false. Thus $f$ is non-decreasing. To see that $f$ is strictly increasing, we assume otherwise that there are $0<r_1<r_2$ such that $f(r_1)=f(r_2)$. Hence
$f$ is constant-valued over $[r_1,r_2]$ which is impossible.

The proof of the lemma is complete.
\end{proof}

\begin{lemma}
For the solution $(a,f,g)$, we have the asymptotic estimates
\begin{equation}\label{decay}
 a(r)={\rm\mbox{O}}\left({\rm\e}^{-\gamma
(1-\varepsilon)r}\right),\quad g(r)=q+{\rm\mbox{O}}\left(r^{-1}\right),\quad f(r)=(\pi-\omega)+{\rm\mbox{O}}\left(r^{-1}\right),\end{equation}
as $r\to\infty$, where $\varepsilon\in(0,1)$ is arbitrarily small and
\be 
\gamma=\frac12\sqrt{\sin^2\omega-2{q^2}}.
\ee
Moreover, the exponential decay rate for $a(r)$ stated in (\ref{decay}) is uniform with respect to the parameter $q$ when $q$ is restricted to any allowed interval
$[0,q_0]$ where $q_0$ satisfies
\be \label{q0}
0<q_0<\min\left\{\frac1{\sqrt{2}}\sin\omega,\sqrt{2}\left(1-\frac\omega\pi\right)\right\}.
\ee
\end{lemma}
\begin{proof} From (\ref{3.17}) and (\ref{3.22}), we see that $f(r)\to\pi-\omega$ uniformly fast for $q\in[0,q_0]$.
Applying this and the other properties derived in Lemma \ref{lemma 6} for $a,f,g$ in the equation (\ref{4}), we have
\be \label{5.3}
a''\geq \left(\frac14\sin^2\omega (1-\delta) -\frac12q^2\right) a,\quad r>R_\delta,
\ee
where $R_\delta>0$ is sufficiently large but independent of $q\in[0,q_0]$ and $\delta>0$ is arbitrarily small. Write
\be 
\frac14\sin^2\omega (1-\delta) -\frac12q^2=\frac14\left(\sin^2\omega  -2q^2\right)(1-\vep)^2,
\ee
and $r_\vep=R_\delta$. Then (\ref{5.3}) gives us $a''\geq\gamma^2(1-\vep)^2 a$, $r>r_\vep$.
Using the comparison function $\sigma (r)=C_\vep\e^{-\gamma(1-\vep)r}$, we have 
that $(a-\sigma)''\geq \gamma^2(1-\vep)^2 (a-\sigma)$, $r>r_\vep$. Thus, by virtue of the maximum principle, we have $(a-\sigma)(r)<0$
for all $r>r_\vep$ when the constant $C_\vep$ is chosen large enough so that $a(r_\vep)\leq \sigma(r_\vep)$. This establishes the uniform exponential decay
estimate for $a(r)$ as $r\to\infty$ with respect to $q\in[0,q_0]$.

To get the estimate for $g$, we note from (\ref{3.12}) that
\be 
g'(r)=\frac1{r^2}\int_0^r 2a^2(\rho) g(\rho)\,\dd\rho,\quad r>0,
\ee
which leads to
\be 
q-g(r)=\int_r^\infty \frac1{\rho^2}\int_0^\rho 2a^2(\rho') g(\rho')\,\dd\rho'\,\dd\rho=\mbox{O}(r^{-1}),
\ee
for $r>0$ large, since $a(r)$ vanishes exponentially fast at $r=\infty$.

To study the asymptotic behavior of $f$, we integrate (\ref{5}) over the interval $(r_0,r)$ ($0<r_0<r<\infty$) to get
\be \label{5.6}
8\kappa \,a^2(r)\sin^2f(r) f'(r)+r^2f'(r)=C_0+\int_{r_0}^r F(\rho)\,\dd r,
\ee
where $C_0$  and $F(r)$ are given by 
\bea 
C_0&=&8\kappa \,a^2(r_0)\sin^2f(r_0) f'(r_0)+r_0^2f'(r_0),\\
F&=&2a^2\sin f \cos f+8\kappa a^2\sin f\cos f(f')^2
+\frac{8\kappa a^4\sin^3f\cos f}{r^2}.
\eea
Take $r_0>0$ large enough so that $\frac12(\pi-\omega)\leq f(r)<\pi-\omega$ for $r\geq r_0$. Thus
\be \label{5.9}
0<\sin\frac12(\pi-\omega)\leq \sin f(r),\quad r\geq r_0.
\ee
Using (\ref{5.9}) and recalling the definition of ${\cal E}_1$, we see that the integral
\be 
\int^\infty_{r_0}a^2\sin f\cos f(f')^2\,\dd r
\ee
is convergent. Applying this result and the exponential decay estimate of $a(r)$ as $r\to\infty$, we obtain
\be \label{5.11}
f'(r)=\frac{C_0+\int_{r_0}^r F(\rho)\,\dd\rho}{8\kappa \,a^2(r)\sin^2f(r)+r^2}=\mbox{O}(r^{-2}),\quad r>r_0.
\ee
Integrating (\ref{5.11}) over $(r,\infty)$ ($r>r_0$), we arrive at
\be 
(\pi-\omega)-f(r)=\mbox{O}(r^{-1}).
\ee

The proof of the lemma is complete.
\end{proof}

\begin{lemma} For the solution $(a,f,g)$ with fixed $\omega\in (\frac\pi2,\pi)$, the electric charge 
\be \label{Qq}
Q_e(q)=2\int_0^\infty a^2(r) g(r)\,\dd r
\ee
enjoys the property $Q_e(q)\to0$ as $q\to0$.
\end{lemma}
\begin{proof} For fixed $\omega$, let $q_0$ satisfy (\ref{q0}). Since $a$ vanishes exponentially fast at infinity uniformly with respect to $q\in (0,q_0]$ and
$0<g(r)<q$ for all $r>0$, we see that we can apply the dominated convergence theorem to (\ref{Qq}) to conclude that $Q_e(q)\to0$ as $q\to0$.
\end{proof}

\medskip 

It should be noted that the special case $\kappa=0$ is to be treated separately since the study above relies on the condition $\kappa>0$ (see (\ref{fn})).
In fact, when $\kappa=0$, the Skyrme term in (\ref{L}) is absent and the model becomes the gauged sigma model which is easier. However, technically, the boundary condition
$f(0)=0$ has to be removed from the definition of the admissible space $\cal A$ but recovered later for the obtained solution to the constrained minimization problem as is done for
the function $g$ in the proof of Lemma \ref{lemma 2}. The details are omitted here.

\small{

}


\begin{thebibliography}{99}

\bibitem{A}
A. Actor, Classical solutions of $SU(2)$ Yang--Mills theories,
{\em Rev. Modern Phys.} {\bf51} (1979) 461--525.

\bibitem{AR}
J. Ambjorn and V. A. Rubakov,
Classical versus semiclassical electroweak decay of a techniskyrmion
{\em Nucl. Phys. } B {\bf256}, 434--448 (1985).

\bibitem{AT}
K. Arthur and  D. H. Tchrakian,
$SO(3)$ gauged soliton of an $O(4)$ sigma model on $\bfR^3$,
{\em Phys. Lett.} B {\bf378} (1996) 187--193. 


\bibitem{BPST}
A. A. Belavin, A. M. Polyakov, A. S. Schwartz, and Yu. S. Tyupkin,
Pseudoparticle solutions of the Yang--Mills equations, {\em Phys.
Lett.} B {\bf59} (1975) 85--87.

\bibitem{BR}
V. Benci and P. H. Rabinowitz, Critical point theorems for indefinite functional, {\em Invent. Math.} {\bf52} (1979) 241--273.

\bibitem{B}
E. B. Bogomol'nyi, The stability of classical solutions, {\em Sov.
J. Nucl. Phys.} {\bf24} (1976) 449--454.


\bibitem{BHT}
Y. Brihaye, B. Hartmann, and D. H. Tchrakian, Monopoles and dyons in
$SO(3)$ gauged Skyrme models, {\em Jour. Math. Phys.} {\bf42} (2001) 3270--3281.

\bibitem{BKT}
Y. Brihaye, B. Kleihaus, and D. H. Tchrakian, Dyon-Skyrmion lumps, {\em J. Math. Phys.}
{\bf40} (1999) 1136--1152.


\bibitem{CW}
C. G. Callan Jr. and E. Witten,
Monopole catalysis of Skyrmion decay, {\em Nucl. Phys.} B {\bf239} (1985)
161--176.

\bibitem{CGSY}
R. Chen, Y. Guo, D. Spirn, and Y. Yang,
Electrically and magnetically charged vortices in the Chern--Simons--Higgs theory,  {\em Proc. Roy. Soc.} A {\bf465} (2009) 3489--3516. 


\bibitem{CM}
Y. M. Cho and D. Maison,
 Monopole configuration in Weinberg--Salam model,
{\em Phys. Lett.} B
{\bf391} (1997)  360--365.

\bibitem{Curie}
P. Curie, On the possible existence of magnetic conductivity and free magnetism, {\em S\'{e}ances Soc.
Phys.} (Paris), pp. 76--77, 1894.

\bibitem{DF}
E. D'Hoker and E. Farhi,
Skyrmions and/in the weak interactions,
{\em Nucl. Phys.} B {\bf241} (1984) 109--128.

\bibitem{D}
P. A. M. Dirac, Quantised singularities in the electromagnetic
field, {\em Proc. Roy. Soc.} A {\bf133} (1931) 60--72.

\bibitem{Eilam}
G. Eilam, D. Klabucar, and A. Stern,
Skyrmion solutions to the Weinberg--Salam model, {\em Phys. Rev. Lett.} 
{\bf56} (1986) 1331--1334.

\bibitem{FR}

D. G. de Figueiredo, J. M. do O, and B. Ruf, 
Critical and subcritical elliptic systems in dimension two, 
{\em Indiana Univ. Math. J.} {\bf 53} (2004) 1037--1054. 


\bibitem{GP}
T. Gisiger, M.B. Paranjape,
Recent mathematical developments in the Skyrme model,
{\em Phys. Reports} {\bf306} (1998) 109--211. 

\bibitem{Go}
P. Goddard and D. I. Olive, Magnetic monopoles in gauge field theories, {\em Rep. Prog. Phys.} {\bf41} (1978) 1357--1437.

\bibitem{Gr}
J. Greensite,
{\em An Introduction to the Confinement Problem},
Lecture Notes in Physics
{\bf821}, Springer-Verlag, Berlin and New York, 2011.

\bibitem{GM}
W. Greiner and J. A. Maruhn, {\em Nuclear Models}, Springer, Berlin, 1996.

\bibitem{JR}
R. Jackiw and C. Rebbi, Degrees of freedom in pseudoparticle
systems, {\em Phys. Lett.} B {\bf67} (1977) 189--192.

\bibitem{JT}
A. Jaffe and C. H. Taubes, {\em Vortices and Monopoles},
Birkh\"{a}user, Boston, 1980.

\bibitem{JZ}
B. Julia and A. Zee, Poles with both electric and electric charges
in non-Abelian gauge theory, {\em Phys. Rev.} D {\bf11} (1975)
2227--1232.

\bibitem{LPY}
C. S. Lin, A. C. Ponce, and Y. Yang,
 A system of elliptic equations arising in Chern--Simons field theory,  {\em J. Funct. Anal.} {\bf 247} (2007)   289--350. 


\bibitem{LY}
F. Lin and Y. Yang, Existence of dyons in the Coupled Georgi--Glashow--Skyrme model, {\em Annales de Henri Poincar\'{e}} {\bf 12} (2011) 329--349. 

\bibitem{MRS}
V. G. Makhankov, Y. P. Rybakov, and V. I. Sanyuk, {\em The
Skyrme Model}, Springer, Berlin and Heidelberg, 1993.

\bibitem{Man}
S. Mandelstam, General introduction to confinement,
{\em Phys. Rep.} C {\bf67} (1980) 109--121.

\bibitem{MS}
N. Manton and P. Sutcliffe, {\em Topological Solitons}, Cambridge Monographs on Mathematical Physics, Cambridge U. Press, Cambridge, 2004. 

\bibitem{PT}
B. Piette and D. H. Tchrakian,
Static solutions in the $U(1)$ gauged Skyrme model,
{\em Phys. Rev.} D {\bf62} (2000) 025020

\bibitem{Po}
 A. M. Polyakov,  Particle spectrum in quantum field theory,  {\em JETP
Lett.} {\bf{20}} (1974) 194.


\bibitem{PS}
M. K. Prasad and C. M. Sommerfeld, Exact classical solutions for
the 't Hooft monopole and the Julia--Zee dyon, {\em Phys. Rev.
Lett.} {\bf35} (1975) 760--762.


\bibitem{R}
R. Rajaraman, {\em Solitons and Instantons}, North-Holland,
Amsterdam, 1982.

\bibitem{Ryder}
L. H. Ryder, {\em Quantum Field Theory}, 2nd ed., Cambridge U. Press, Cambridge, U. K., 1996.


\bibitem{SW}
M. Schechter and R. Weder,  A theorem on the existence of dyon
solutions, {\em Ann. Phys.} {\bf{132}} (1981) 293--327.


\bibitem{S}
J. Schwinger, A magnetic model of matter, {\em Science} {\bf165}
(1969) 757--761.

\bibitem{SY}
M. Shifman and A. Yung,
{\em Supersymmetric Solitons}, Cambridge U. Press, Cambridge, U. K., 2009.

\bibitem{S1}
T. H. R. Skyrme, A unified field theory of mesons and baryons, {\em Nucl. Phys.} {\bf31} (1962) 556--569.

\bibitem{S2}
T. H. R. Skyrme, The origins of Skyrmions, {\em Internat. J. Mod. Phys.} A {\bf3} (1988) 2745--2751.



\bibitem{Taubes}
C. H. Taubes, The existence of a non-minimal solution to the
$SU(2)$ Yang--Mills--Higgs equations on $\bf R^3$, Parts I and II,
{\em Commun. Math. Phys.} {\bf86} (1982) 257--320.


\bibitem{'t}
G. 't Hooft, Magnetic monopoles in unified gauge theories, {\em Nucl. Phys.} B {\bf79} (1974) 276--284.

\bibitem{tH0}
G. 't Hooft, Computation of the quantum effects due to a four-dimensional pseudoparticle, {\em Phys. Rev.} D {\bf14} (1976) 3432--3450.

\bibitem{tH}
G. 't Hooft, A property of electric and magnetic flux in
nonabelian gauge theories, {\em Nucl. Phys.} B {\bf 153} (1979)
141--160.

\bibitem{W1}
E. Witten,
Global aspects of current algebra,
{\em Nucl. Phys.} B {\bf223} (1983) 422--432.

\bibitem{Yw}
Y. Yang,
Dually charged particle-like solutions in the Weinberg--Salam theory, {\em Proc. Roy. Soc.} A  {\bf454} (1998) 155--178.


\bibitem{Ybook}
Y. Yang, {\em Solitons in Field Theory and Nonlinear Analysis},
Springer, New York, 2001.

\bibitem{ZB}
I. Zahed and G. E. Brown, The Skyrme model, {\em Phys.
Reports} {\bf142} (1986) 1--102.

\end{thebibliography}
\end{document}